\documentclass[sigconf]{acmart}

\usepackage{graphicx}
\usepackage{amsmath}

\usepackage{amssymb}
\usepackage{booktabs}
\usepackage{amsfonts}
\usepackage{algorithm}
\usepackage{algorithmicx}
\usepackage{algpseudocode}
\usepackage{footnote}

\usepackage{xcolor}
\usepackage{comment}
\usepackage{amsthm}
\usepackage{bbm} 
\usepackage{subcaption}
\usepackage{multirow}
\usepackage{comment}
\usepackage{makecell} 
\usepackage{tabularx}
\usepackage[normalem]{ulem}
\useunder{\uline}{\ul}{}
\usepackage{booktabs}
\usepackage[switch]{lineno}

\usepackage{url}
\usepackage{xspace}

\usepackage[export]{adjustbox}
\usepackage{paralist} 

\usepackage{tabu}

\newcommand{\nbf}[1]{{\noindent \textbf{#1}}}

\newcommand{\eat}[1]{}

\newcommand{\cbit}{\begin{compactitem}}
\newcommand{\ceit}{\end{compactitem}}
\newcommand{\cben}{\begin{compactenum}}
\newcommand{\ceen}{\end{compactenum}}

\usepackage[capitalize]{cleveref}
\crefname{section}{Sec.}{Secs.}
\crefname{table}{Table}{Tables}
\crefname{figure}{Fig.}{Figs.}
\crefname{algocf}{alg.}{algs.}
\Crefname{algocf}{Algorithm}{Algorithms}

\AtBeginDocument{%
  }

\newcommand{\benchname}{\texttt{MTRB}\xspace}
\newcommand{\methodname}{QTA\xspace}

\definecolor{customred}{rgb}{0.95, 0.35, 0.45}
\definecolor{customgreen}{rgb}{0.35, 0.75, 0.5}
\definecolor{customblue}{rgb}{0.35, 0.45, 0.95}
\definecolor{customgolden}{HTML}{FDD769}
\definecolor{custompurple}{HTML}{9999FF}

\def\paperTitle{
Data-Efficient Massive Tool Retrieval: A Reinforcement Learning Approach for Query-Tool Alignment with Language Models
}

\def\authorBlock{
\author{Yuxiang Zhang}
\authornote{Authors contributed equally to this research.}
\affiliation{
  \institution{Waseda University}
  \city{Tokyo}
  \country{Japan}
}
\email{joel0495@asagi.waseda.jp}

\author{Xin Fan}
\authornotemark[1]
\affiliation{
  \institution{Waseda University}
  \city{Tokyo}
  \country{Japan}
}
\email{fan_xin@fuji.waseda.jp}

\author{Junjie Wang}
\authornotemark[1]
\affiliation{
  \institution{Waseda University}
  \city{Tokyo}
  \country{Japan}
}
\email{wjj1020181822@toki.waseda.jp}

\author{Chongxian Chen}
\affiliation{
  \institution{Waseda University}
  \city{Tokyo}
  \country{Japan}
}
\email{chenc@toki.waseda.jp}

\author{Fan Mo}
\affiliation{
  \institution{Waseda University}
  \city{Tokyo}
  \country{Japan}
}
\email{bakubonn@toki.waseda.jp}

\author{Tetsuya Sakai}
\authornote{Corresponding Author}
\affiliation{
  \institution{Waseda University}
  \city{Tokyo}
  \country{Japan}
}
\email{tetsuyasakai@acm.org}

\author{Hayato Yamana}
\authornotemark[2]
\affiliation{
  \institution{Waseda University}
  \city{Tokyo}
  \country{Japan}
}
\email{yamana@yama.info.waseda.ac.jp}
}

\copyrightyear{2024}
\acmYear{2024}
\setcopyright{acmlicensed}
\acmConference[SIGIR-AP '24] {Proceedings of the 2024 Annual International ACM SIGIR Conference on Research and Development in Information Retrieval in the Asia Pacific Region}{December 9--12, 2024}{Tokyo, Japan.}
\acmBooktitle{Proceedings of the 2024 Annual International ACM SIGIR Conference on Research and Development in Information Retrieval in the Asia Pacific Region (SIGIR-AP '24), December 9--12, 2024, Tokyo, Japan}
\acmISBN{979-8-4007-0436-9/24/10}
\acmDOI{10.1145/XXXXXX.XXXXXX}

\begin{document}

\title{\paperTitle}

\authorBlock

\begin{abstract}
Recent advancements in large language models (LLMs) integrated with external tools and APIs have successfully addressed complex tasks by using in-context learning or fine-tuning.
Despite this progress, the vast scale of tool retrieval remains challenging due to stringent input length constraints. 
In response, we propose a pre-retrieval strategy from an extensive repository, effectively framing the problem as the massive tool retrieval (MTR) task. 
We introduce the \benchname (massive tool retrieval benchmark) to evaluate real-world tool-augmented LLM scenarios with a large number of tools.
This benchmark is designed for low-resource scenarios and includes a diverse collection of tools with descriptions refined for consistency and clarity. It consists of three subsets, each containing $90$ test samples and $10$ training samples.
To handle the low-resource MTR task, we raise a new query-tool alignment (\methodname) framework leverages LLMs to enhance query-tool alignment by rewriting user queries through ranking functions and the direct preference optimization (DPO) method.
This approach consistently outperforms existing state-of-the-art models in top-$5$ and top-$10$ retrieval tasks across the \benchname benchmark, with improvements up to $93.28\%$ based on the metric Sufficiency@$k$, which measures the adequacy of tool retrieval within the first $k$ results. 
Furthermore, ablation studies validate the efficacy of our framework, highlighting its capacity to optimize performance even with limited annotated samples. 
Specifically, our framework achieves up to $78.53\%$ performance improvement in Sufficiency@$k$ with just a single annotated sample.
Additionally, \methodname exhibits strong cross-dataset generalizability, emphasizing  its potential for real-world applications.
\end{abstract}



\maketitle

\section{Introduction}

\begin{figure}[!tp]
\centering
\includegraphics[width=0.47\textwidth]{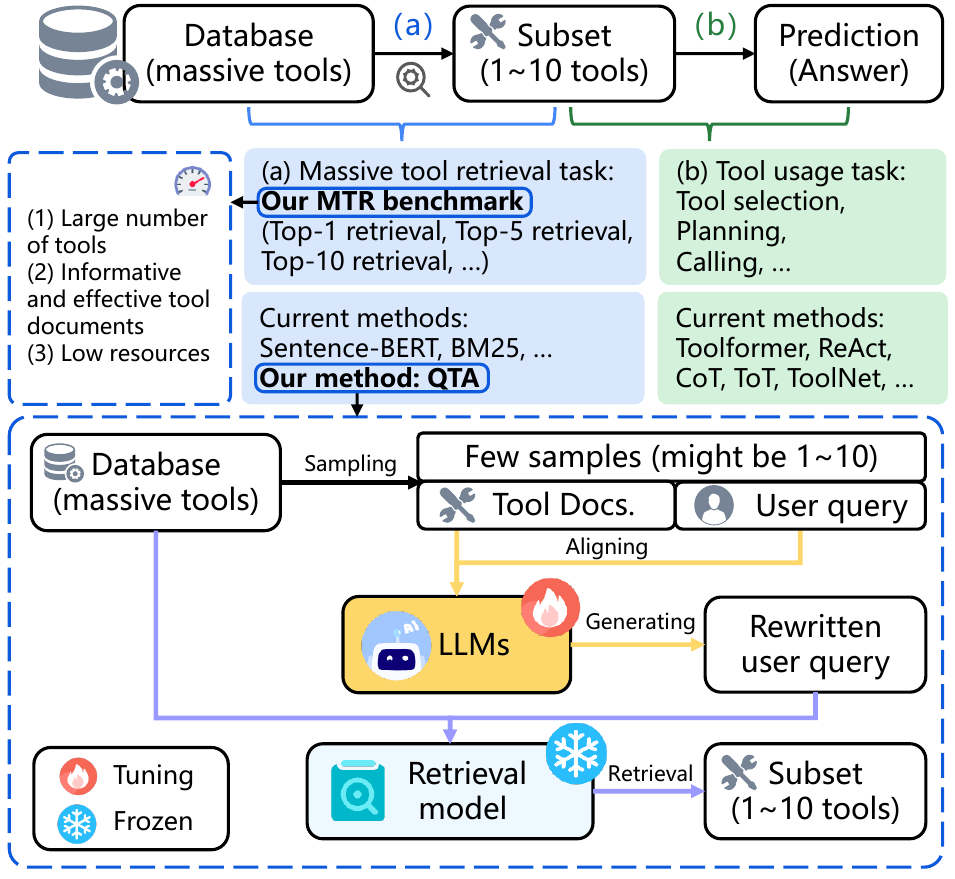}
\caption{The current approach to solving tool-based problems involves first addressing the (a) massive tool retrieval (MTR) task, followed by completing the (b) tool selection task. We focus on providing a solution for the MTR task. For evaluation, we introduce a new \benchname benchmark. Methodologically, we propose a new \methodname framework to enhance the retrieval systems by aligning user queries with tools.}
\label{fig:introduction}
\vspace{-1em}
\end{figure}

Recent advancements show tool-equipped large language models (LLMs) can effectively handle various complex tasks, including mathematical problems~\cite{DBLP:conf/nips/HaoLWH23@toolkengpt,DBLP:journals/corr/abs-2302-07842@aug-survey}.
Specifically, several studies have indicated that in-context learning or fine-tuning methods offer considerable potential for solving tool usage problems~\cite{DBLP:conf/nips/LiuTMMHBR22@intro-icl-ft1, DBLP:journals/corr/abs-2310-03331@intro-icl-ft2, DBLP:conf/acl/WuWY0FXQ23@intro-icl-ft3}. 
For example, API-Bank~\cite{DBLP:conf/emnlp/LiZ000YLHL23@apibank} can select the best tool from a small set of tools for a specific task and then engage with it interactively.
However, as society progresses, a vast array of tools emerge, with some real-world applications encompassing thousands of tools. 
Unfortunately, current methods struggle with these tasks based on large-scale tools due to inherent model design constraints. 
For example, the maximum input length for Llama-2 series models~\cite{DBLP:journals/corr/abs-2307-09288@llama2} is $4096$ characters. However, inputting information about a thousand tools, including their names and descriptive documents, into the model could require over $100,000$ characters~\cite{DBLP:journals/corr/abs-2401-06201@easytool}, far beyond the models' context capacity.
To mitigate this challenge, recent research has introduced a method where tools are actively screened through a retrieval system before being fed into LLMs. 
The retrieval system selectively identifies the top $k$ tools that match a user's query, forming a targeted subset. 
Consequently, as shown in~\cref{fig:introduction} (a), this task can be redefined as the \textit{massive tool retrieval (MTR)} task.

However, current benchmarks are not designed with the MTR task in mind and only consider tool usage problems, such as calling and planning. 
To comprehensively evaluate the capabilities of retrieval systems in the MTR task, we propose the \benchname benchmark adhering to three criteria: (1) a large number of tools, (2) informative and effective tool documents, and (3) low resources.
Specifically, we reorganize the tools within the RestBench~\cite{DBLP:journals/corr/abs-2306-06624@restbench}, MetaTool~\cite{DBLP:journals/corr/abs-2310-03128@metatool}, and ToolBench~\cite{DBLP:journals/corr/abs-2307-16789@toolbench} datasets, encompassing a diverse array of $2,645$ tools from various domains, such as weather and music. 
Furthermore, we developed a text optimization workflow that enhances the original tool descriptions into informative and effective documents. 
To address the low-resource criterion, we collect a total of $300$ samples and only use $10\%$ ($30$ samples) for training, with the remaining $270$ samples serving as the test set. 
This setup allows us to validate the retrieval systems under low-resource conditions.

As shown in~\cref{fig:introduction}, current tool-based research predominantly focuses on tool usage problems while neglecting MTR tasks.
Few studies have explored potent retrieval-based methods.
Specifically, these studies utilize fine-tuned models like Sentence-BERT~\cite{DBLP:conf/emnlp/ReimersG19@sentencebert} for feature extraction, followed by ranking based on cosine similarity.
However, these methods necessitate extensive training samples, subsequently requiring lots of new manually annotated data for task transfer. 
Specifically, we note that tool documents in real-world applications and our \benchname benchmark are rich in information and formatted. 
However, the diversity in user queries can introduce biases in the similarity calculations of retrieval models. 
To mitigate this, as shown in~\cref{fig:introduction}, we propose a low-resource effective query-tool alignment (\methodname) framework that leverages reinforcement learning and utilizes hidden ranking information within MTR tasks. 
Specifically, we employ LLMs to re-write user queries by understanding tool documents and user intentions. 
Our framework also introduces a new ranking function to optimally rank these rewritten queries. 
Consequently, the \methodname framework aligns closely with user preferences and tool documentation. 
By utilizing the direct preference optimization (DPO) training~\cite{DBLP:conf/nips/RafailovSMMEF23@dpo}, our framework achieves effective training with minimal annotated data, even with only one sample, and demonstrates strong transfer-ability to new tasks (Details in~\cref{ss:training}).
  
We conduct extensive experiments on the \benchname benchmark across multiple subsets, showing that our \methodname framework consistently achieves state-of-the-art (SOTA) performance. 
Specifically, in the Sufficiency@$k$ retrieval tasks, which assess the inclusion of all necessary tools within the top $k$ results, our method significantly outperforms the baseline. 
On the \benchname-RestBench subset, \methodname improves the Sufficiency@$5$ performance by $93.28\%$ relative to the baseline method, increasing the score from $16.67$ to $32.22$.
Moreover, our ablation studies validate the robustness of our design and the efficacy of our data usage. 
For instance, with just one annotated sample, our \methodname achieves a $78.53\%$ Sufficiency@$5$ improvement on the \benchname-RestBench subset, effectively demonstrating its capability to ensure comprehensive tool availability for task execution.

In summary, our contributions are as follows.
\begin{itemize}
\item We introduce a \benchname benchmark for evaluating the MTR task, including a wide array of tools and their associated informative and effective documents. Focus on massive tool counts and low-resource scenarios.
\item To address the MTR task, we introduce \methodname, a new data-efficient alignment framework that leverages reinforcement learning and utilizes hidden ranking information in MTR tasks. Our method achieves up to $78.53\%$ Sufficiency@$5$ improvements using just a single annotated sample, underscoring its efficiency in leveraging scarce labeled data.
\item We conduct extensive experiments to establish the baselines for our benchmark and to validate the effectiveness of the proposed framework. The experimental results demonstrate that our benchmark presents a challenge to existing retrieval systems, and show that the proposed framework substantially improves the performance of baselines.
\end{itemize}

\section{Related Work}

\subsection{Massive Tool Retrieval v.s. Tool Selection}

Tool selection (TS) tasks require LLMs to identify the appropriate tools from a small set of candidates based on user queries. 
As illustrated in~\cref{fig:introduction} (b), the aim of this task, like other tool usage tasks, is to assess the understanding and reasoning capabilities of LLMs in using tools. 
For example, in the ToolBench~\cite{DBLP:journals/corr/abs-2307-16789@toolbench}, the TS task involves selecting one tool from $5$ or $6$ candidate tools. 
The MetalTool dataset~\cite{DBLP:journals/corr/abs-2310-03128@metatool} requires selecting multiple tools from $10$ candidates.
However, as the number of tools increases rapidly, we must manage a substantial tool repository, not merely a small collection. 
Consequently, some research~\cite{DBLP:journals/corr/abs-2310-03128@metatool,DBLP:journals/corr/abs-2307-16789@toolbench} suggests employing a retrieval model for the preliminary filtering of this extensive set.
Building on this, we standardize the massive tool retrieval (MTR) task. 
The goal of the MTR task is to retrieve an optimal subset of tools from a vast tool repository in response to a user query. 
This subset aims to be as small as possible while including the necessary tools for addressing downstream tool usage tasks.
Therefore, to comprehensively assess the performance of the retrieval system on the MTR task, we introduce a new \benchname benchmark. 
Specifically, we drew on existing benchmarks that include TS tasks and the methodologies used in various retrieval scenarios, such as article retrieval, while considering an extensive array of tools and comprehensive tool documentation~\cite{DBLP:journals/corr/abs-2401-06201@easytool,DBLP:journals/ir/QinLXL10@letor,DBLP:conf/nips/WangWWNBP23@doris-mae}.

\subsection{Tool Retrieval Systems}

To facilitate the MTR task, several fine-tuning techniques are employed to develop models capable of calculating the similarity between tools and user queries.
For example, Toolbench fine-tunes a BERT model on more than $43$k annotated samples, enabling efficient retrieval from a database containing over $16$k tools~\cite{DBLP:journals/corr/abs-2307-16789@toolbench}.
However, these methods lack transferability, necessitating extensive manual annotation and fine-tuning for unseen datasets.
To address this challenge, models specifically designed for document retrieval have been pre-trained on a vast corpus of annotated data and subsequently transferred to retrieval tasks by fine-tuning. 
This strategy offers a promising pathway for addressing the MTR task.
For instance, the Sentence-BERT model~\cite{DBLP:conf/emnlp/ReimersG19@sentencebert} is fine-tuned on over $1$ million annotated sentence pairs utilizing a 3-way softmax classifier objective function, while the all-MiniLM-L6-v2 model~\cite{DBLP:conf/nips/WangW0B0020@minilm} has been fine-tuned on over $1$ billion sentence pairs. 
However, these approaches encounter challenges in MTR tasks, particularly when user queries are based on specific tool usage tasks. 
For example, a request such as ``give me a movie cover from the Harry Potter collection'' requires the integration of multiple tool actions, not directly linked to a single request. 
This involves coordinating several tools, such as ``GET /search/collection'', ``GET /collection/\{collection\_id\}'', and ``GET /movie/\{movie\_id\}/images'', to fulfill the user's request.
These methods exhibit such gaps in training set design, particularly in the construction of similar sentence pairs that only partially mirror real-world scenarios. 
Therefore, to mitigate this challenge, we propose the \methodname framework to align user queries with tool documents. 
Moreover, only a few annotated data samples are needed to train the framework.

\section{\benchname Benchmark}
In this section, we introduce the MTR task. 
Subsequently, we systematically create our \benchname benchmark, incorporating procedures such as employing a text optimization workflow.

\subsection{Task Formulation}
Inspired by previous research~\cite{DBLP:journals/corr/abs-2307-16789@toolbench,DBLP:journals/corr/abs-2310-03128@metatool}, the MTR task aims to retrieve the necessary tools from an extensive tool database ($T$) containing $M$ tools, based on the user query ($q$). 
In detail, the output is a small subset of essential tools, denoted as $GT$ (short for ``Golden Tools''), consisting of $N$ items deemed most relevant. 
Each tool in the database is defined by its name and related documents.
The process can be shown as:
\begin{equation}
\mathcal{R}(q, T) = GT \subset T.
\end{equation}

Specifically, $T$ typically contains a large number of tools, ranging from dozens to thousands.

\subsection{Data Curation}
To establish a robust benchmark, we adhere to three core principles: (1) a large number of tools; (2) informative and effective tool documents; (3) low resource scenarios.
Based on these principles, we develop the following procedure.

\nbf{Step 1: Tool extraction.} 
We extract all tools in three widely-used large-scale tool-based datasets (RestBench~\cite{DBLP:journals/corr/abs-2306-06624@restbench}, MetaTool~\cite{DBLP:journals/corr/abs-2310-03128@metatool}, ToolBench~\cite{DBLP:journals/corr/abs-2307-16789@toolbench}), including tool names, documents, and other related information. 

\nbf{Step 2: Text optimization workflow.}
We observe several issues with the tool documents, including incomplete or invalid information (such as errors, blanks, or corrupted entries), redundancy, and considerable length disparities. 
To address these problems, considering some optimization strategies~\cite{DBLP:journals/corr/abs-2401-06201@easytool}, we undertake a systematic text optimization of the documents.
In detail, we manually process documents by referring to the original information, implementing modifications such as abbreviations, expansions, and re-writings. 
Furthermore, to address the disparity in token numbers~\cite{DBLP:conf/acl/TanSWYS22@embedding-length1} and enhance readability, we manually restrict the length of tool documents to a range near the median count of original tokens.

\nbf{Step 3: Sample preparation.} 
Inspired by the designs of tool selection tasks in benchmarks~\cite{DBLP:journals/corr/abs-2307-16789@toolbench,DBLP:journals/corr/abs-2310-03128@metatool}, we extract question-answer pairs from three datasets, focusing solely on user queries and tool names. 
To ensure the reliability of our data, we conduct manual validations of these samples, preventing any mismatches between queries and answers.

\nbf{Step 4: Filtering.}
All samples are subjected to a manual review process aimed at identifying and eliminating any content deemed inappropriate, including material with clear ethical concerns.

After the aforementioned steps, we retain a total of $300$ samples, including $30$ training samples and $270$ test samples. 
Each subset contains $10$ training samples and $90$ test samples. 
Furthermore, we provide a general statistical overview in~\cref{tab:key_statistic}. 
In detail, we utilize the Llama-3 tokenizer~\cite{metaAI@llama3} to tokenize the text, facilitating the length statistics. 
We consider key information, such as the total number of tools and the range of token lengths in the tool documents.

\begin{table}[tp]
\setlength{\tabcolsep}{3pt}
\small
\centering
\caption{General statistics of \benchname benchmark.
Tool Doc. Lengths represent the length range of tool descriptions. 
The Golden Tools column indicates the number of essential tools selected as ground truth for each sub-task.}
\label{tab:key_statistic}
\begin{tabular}{l|c|c|c}
\toprule
Sub-task & \# of Tools & Tool Doc. Lengths & \# of Golden Tools  \\
\midrule
\benchname-RestBench & 54 & 20-30 tokens & \{1, 2, 3, 4\} \\
\midrule
\benchname-MetaTool  & 199 & 10-20 tokens & \{1\} \\
\midrule
\benchname-ToolBench & 2391 & 70-100 tokens & \{2, 3\} \\
\bottomrule
\end{tabular}
\end{table}

\subsection{Evaluation Metrics}
Given that our task is a retrieval task, we employ widely-used metrics such as Recall@$k$ and NDCG@$k$~\cite{DBLP:journals/tois/JarvelinK02@ndcg}. 
Specifically, Recall@$k$ measures the ratio of golden tools retrieved within the top $k$ results to the total number of golden tools. 
In tool-augmented LLM scenarios, it is necessary to provide the LLM with all the requisite tools to complete the task, rather than just a subset of the tools. 
To characterize this, we propose the Sufficiency@$k$. We define Sufficiency@$k$ as a binary metric; it yields a value of $1$ if the set of tools retrieved is deemed adequate for task completion, and a value of $0$ otherwise.
This metric gauges whether the first $k$ retrieval results include all the tools needed to complete the task, which is crucial for ensuring that the LLM can successfully execute complex tasks. 
Beyond the Recall@$k$ metric, which would still assign some points even if the retrieved results only included some of the required tools, the Sufficiency@$k$ metric places greater emphasis on the completeness of the retrieved results.
Moreover, NDCG@$k$ (Normalized Discounted Cumulative Gain at $k$) considers the relevance of the tools and their positions in the results list. 
The core idea is that tools appearing earlier in the results are more important to users than those appearing later.
As shown in~\cref{tab:key_statistic}, the range of golden answers in our benchmark spans from one to four. 
Therefore, we establish two evaluation gradients: top-$5$ and top-$10$. 
Overall, our metrics include Sufficiency@$5$ ($S@5$), Sufficiency@$10$ ($S@10$), NDCG@$5$ ($N@5$), and NDCG@$10$ ($N@10$).

\section{\methodname Framework}
In this section, we outline the proposed query-tool alignment (\methodname) framework.
In detail, we introduce the modules and the process of aligning user queries with tool documents.

\begin{figure*}[!tp]
\centering
\includegraphics[width=0.95\textwidth]{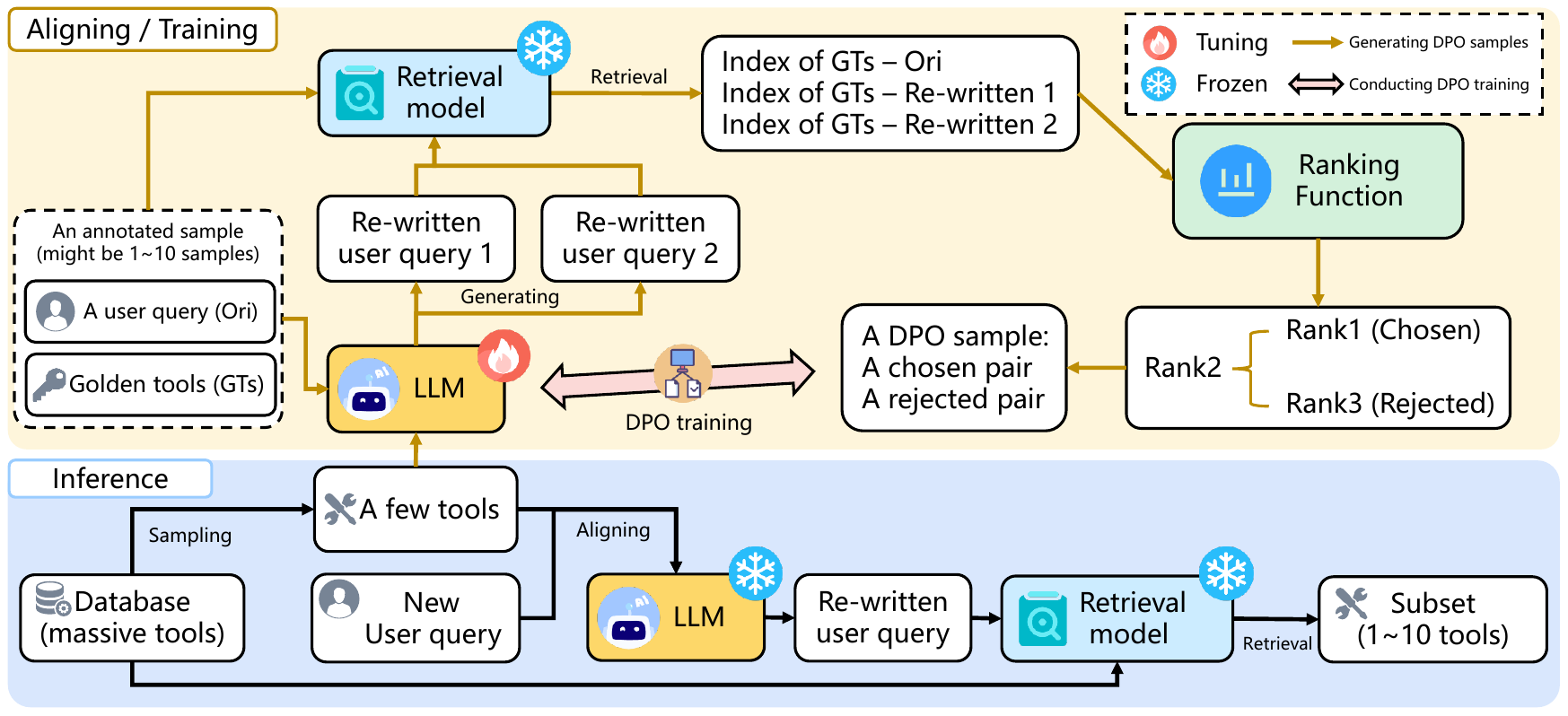}
\caption{An overview of the proposed \methodname framework, which includes data pipelines for the training and inference stages.
Specifically, we utilize an LLM to learn the alignment between user queries and tool document representations, thereby generating high-quality user queries. 
Additionally, we employ a frozen retrieval model to compute the similarity between the queries and the tool database.
}
\label{fig:framework}
\end{figure*}

\subsection{Architecture Overview}
\label{ss:overview}

We propose a framework that leverages the generalization capabilities of LLMs to enhance retrieval models, thereby improving the retrieval effectiveness of tool documentation.
Our architecture consists of two core components: a powerful LLM and a frozen retrieval model. 
Previous research primarily focuses on fine-tuning and optimizing retrieval models~\cite{DBLP:journals/corr/abs-2307-16789@toolbench}. 
However, these methods often require a large number of annotated samples and exhibit weak generalization capabilities when dealing with tool documents that have content variations. 
To overcome these challenges, we utilize an LLM to learn the inherent characteristics of tool databases, thereby aligning user queries with the corresponding tool documentation.
Specifically, to harness the ranking information from retrieval system, we convert these into DPO pairs, which serve as additional training data for the LLM. 
Through reinforcement learning, we optimize the LLM's understanding with minimal annotated samples, bridging the semantic gap between queries and tool documents effectively.

As shown in~\cref{fig:framework}, given a user query ($q$) and a tool database ($T$), we prompt LLM to understand the text from all documents for transforming $q$ into a revised version ($q^{re}$). 
The transformation is shown by the equation:
\begin{equation}
LLM(q, T) = q^{re}.
\label{eq:main_ori}
\end{equation}

However, since the tool database ($T$) is typically extensive and exceeds the input context of LLM, we perform random sampling to obtain a subset containing $s$ tools ($T_{sub} = \{ T_1, T_2, ..., T_s \}$).
This subset fits within the context limitations of the LLM. 
Consequently, \cref{eq:main_ori} is rewritten as:
\begin{equation}
LLM(q, T_{sub}) = q^{re}.
\label{eq:main_sub}
\end{equation}

In this context, we can employ various sampling techniques to generate $T_{sub}$, such as random sampling and tool documents highly relevant to the user query. 
Given our intention to adapt this framework to diverse environments, the random sampling method is adopted as the standard approach.

\subsection{Aligning User Query with Tool Documents}
\label{ss:training}

During the training phase, our goal is to train an LLM to generate high-quality rewritten user query ($q^{re}$), aligning user queries with tool documentation. 
Specifically, we employ the method of direct preference optimization (DPO)~\cite{DBLP:conf/nips/RafailovSMMEF23@dpo} for reinforcement learning, which consists of two main steps: \textbf{generating DPO samples} and \textbf{conducting DPO training}.
For the reinforcement learning algorithm, we select DPO as our primary learning strategy for two considerations: flexibility in data construction, and efficiency in low-resource scenarios.
Unlike other algorithms, such as PPO~\cite{DBLP:journals/corr/SchulmanWDRK17@ppo}, DPO does not require a reward model or manual annotations for the reward model, thereby reducing reliance on computational and human resources. 
Moreover, DPO allows us to start with a few annotated data and iteratively generate preference data through multiple generations, substantially expanding the scale of training samples.
Subsequently, as shown in~\cref{fig:framework}, we explain the training process by presenting a specific annotated sample, which includes an original user query ($q$) and a golden tool set with one golden tool ($GT = \{gt_1\}$).

\nbf{Generating DPO samples.}
A DPO sample requires two pairs: a ``chosen'' pair to guide the LLM towards the desired generation direction, and a ``rejected'' pair to train the LLM to avoid specific outputs.
Therefore, our goal in this step is to obtain these pairs.

Initially, we instruct the LLM to generate two independent rewritten user queries ($\{q^{re}_1, q^{re}_2\}$), following the method described in~\cref{ss:overview}. 
To accelerate training by maximizing the diversity of the rewritten texts, we set a high temperature for the LLM.
Subsequently, considering the need to adapt the LLM's generated texts for the retrieval model, we employ this system to generate indices. 
These indices are references for selecting ``chosen'' and ``rejected'' pairs. 
Specifically, we compute the similarity between the original user query ($q$), the two rewritten user queries ($\{q^{re}_1, q^{re}_2\}$), and the tool database ($T$), recording the indices of the corresponding golden tool set ($GT =\{gt_1\}$).

Furthermore, we introduce a ranking function to score and sort these queries ($\{q, q^{re}_1, q^{re}_2\}$).
The objective of the ranking function is to utilize indices generated by the retrieval model to promote the creation of queries that favor golden tool documents. 
Specifically, we highlight the importance of items ranked within the top $n$ positions and apply escalating penalties to items failing to achieve a top $n$ ranking.
We utilize the widely adopted discounted cumulative gain (DCG) function, which emphasizes top-ranked results by applying a diminishing function to the ranking scores. 
To better align with our task requirements, we introduce specific modifications to the standard DCG approach.
We present an example where the original user query is processed by the retrieval model, resulting in the ranking ($idx$) of the golden tool ($\{gt_1\}$) in $T$. 
The quality of the original user query can be evaluated using the following formula:

\begin{equation}
\text{score}(idx) = 
\begin{cases} 
\frac{1}{\log_2(idx + 1.1)} & \text{if } idx \leq n \\
-\frac{idx - n}{\log_2(\frac{idx}{n} + 1)} & \text{if } idx > n
\end{cases}
\label{eq:ranking-function}
\end{equation}

Specifically, we assign higher rewards than the original DCG function for cases where $idx \leq n$. 
Additionally, for cases where $idx > n$,  we create a score with a negative value, with penalties increasing as the rank goes higher. 
This design aims to reinforce the penalties for rankings that do not make it into the top $n$, thereby encouraging the optimization algorithm to push more elements into the top $n$.
If multiple golden tools are available, we sum their scores to compute the final score for each query.

We apply similar operations to $\{q^{re}_1, q^{re}_2\}$, resulting in $\{idx^{re}_1, idx^{re}_2\}$. 
Subsequently, we rank the queries based on these scores, identifying the top-ranked queries as the ``chosen'' pair and the lower-ranked ones as the ``rejected'' pair.
For instance, if the final results satisfy $idx^{re}_1 < idx < idx^{re}_2$, we designate $\{q, q^{re}_1\}$ as the ``chosen'' pair and $\{q, q^{re}_2\}$ as the ``rejected'' pair. 
This approach yields a DPO sample comprising one ``chosen'' pair and one ``rejected'' pair.

\nbf{Conducting DPO training.} 
After obtaining a DPO sample, the LLM undergoes training using the DPO method.
The core principle of the DPO is to optimize actions based on direct user preferences. 
Specifically, the DPO algorithm employs an optimization objective as follows:

\begin{equation}
\max_{\pi_\theta} \mathbb{E}_{q \sim D, q^{re}_i \sim \pi_\theta(\cdot | q)} [r(q, q^{re}_i)] - \beta D_{KL} [\pi_\theta(\cdot | q) || \pi_{\text{ref}}(\cdot | q)],
\label{eq:dpo-objective}
\end{equation}
where, $\mathbb{E}$ is the expectation over samples;
$D$ represents sampling from the user query ranking dataset;
The $r(q, q^{re}_i)$ is the reward function defined by the inverse KL divergence between the policy $\pi_\theta$ and the reference policy $\pi_{\text{ref}}$;
The $\beta$ serves as a tuning mechanism, balancing the reward against the deviation from the reference policy.

In detail, the reward function \(r(q, q^{re}_i)\) is defined as follows:

\begin{equation} 
r(q, q^{re}_i) = \beta \log \frac{\pi_\theta(q^{re}_i|q)}{\pi_{\text{ref}}(q^{re}_i|q)} + \beta \log Z(q) ,
\label{eq:dpo-reward}
\end{equation}
where, \(Z(q)\) is the normalization factor, ensuring the correctness of the probability distribution; 
The term $q^{re}_i$ represents a variant among the re-written user queries, specifically one of $\{q^{re}_1, q^{re}_2\}$.
By modifying this reward function to align with actual ground-truth rewards, the formula simplifies to:

\begin{equation}
r^*(q, q^{re}_i) = \beta \log \frac{\pi^*(q^{re}_i|q)}{\pi_{\text{ref}}(q^{re}_i|q)} .
\label{eq:dpo-reward*}
\end{equation}

Notice that $Z(q)$ is excluded here since the focus shifts from the learned policy $\pi_{\theta}$ to the optimal policy $\pi^*$.

After that, based on the ``chosen'' pair $\{q, q^{re}_1\}$ and the ``rejected'' pair $\{q, q^{re}_2\}$, we apply the Bradley-Terry ($BT$)~\cite{19ff28b9-64f9-3656-ba40-08326a05748e@bradleymodel} model.
This model employs the human preference distribution $p^*_{BT}$ for pairwise comparisons.
It calculates the probability that the chosen query $q^{re}_1$ is preferred over the rejected query $q^{re}_2$, as described by the equation:

\begin{equation}
p^*_{BT}(q^{re}_1 \succ q^{re}_2 | q) = \sigma(r^*(q, q^{re}_1) - r^*(q, q^{re}_2)),
\label{eq:dpo-bradley-r*}
\end{equation}
where $\sigma$ represents the logistic function, defined as:
\begin{equation}
\sigma(z) = \frac{1}{1 + \exp(-z)}.
\end{equation}

Upon substituting the expressions for $r^*(q, q^{re}_1)$ and $r^*(q, q^{re}_2)$ into the Bradley-Terry model, we derive the following probability:

\begin{equation}
p^*_{BT}(q^{re}_1 \succ q^{re}_2 | q) = \frac{1}{1 + \exp\left(\beta \log \frac{\pi^*(q^{re}_2|q)}{\pi_{\text{ref}}(q^{re}_2|q)} - \beta \log \frac{\pi^*(q^{re}_1|q)}{\pi_{\text{ref}}(q^{re}_1|q)})\right)}.
\label{eq:dpo-bradley}
\end{equation}

This expression indicates that if \( q^{re}_1 \) is more likely to be chosen over \( q^{re}_2 \), the ratio of \( \pi^*(q^{re}_1|q) \) to \( \pi_{\text{ref}}(q^{re}_1|q) \) should be greater than the ratio of \( \pi^*(q^{re}_2|q) \) to \( \pi_{\text{ref}}(q^{re}_2|q) \). To align this with the standard logistic function form, the equation is rewritten as:

\begin{equation}
p^*_{BT}(q^{re}_1 \succ q^{re}_2 | q) = \sigma\left(\beta \log \frac{\pi^*(q^{re}_1|q)}{\pi_{\text{ref}}(q^{re}_1|q)} - \beta \log \frac{\pi^*(q^{re}_2|q)}{\pi_{\text{ref}}(q^{re}_2|q)}\right) .
\label{eq:dpo-bradley-simple}
\end{equation}

DPO then optimizes the Bradley-Terry model probability $p^*_{BT}$ using a negative log-likelihood loss.
For clarity and conciseness in the model's loss function, we define an auxiliary function $z_{\theta}(q, q^{re}_1, q^{re}_2)$ as:

\begin{equation}
z_{\theta}(q, q^{re}_1, q^{re}_2) = \beta \log \frac{\pi^*(q^{re}_1|q)}{\pi_{\text{ref}}(q^{re}_1|q)} - \beta \log \frac{\pi^*(q^{re}_2|q)}{\pi_{\text{ref}}(q^{re}_2|q)} .
\label{eq:dpo-auxiliary}
\end{equation}

This allows us to succinctly represent the loss function as:

\begin{equation}
L_{DPO}(\pi_\theta; \pi_{\text{ref}}) = -\mathbb{E}_{(q, q^{re_1}, q^{re_2}) \sim D} \left[ \log \sigma ( z_{\theta}(q, q^{re}_1, q^{re}_2) ) \right] .
\label{eq:dpo-loss}
\end{equation}

\begin{table*}[!tp]
\setlength{\tabcolsep}{3pt}
\centering
\caption{Main results of \benchname benchmark, with the best score \textbf{bolded} and the second best scores \underline{underlined}. The results are shown in percentages.}
\label{tab:main_results}
\vspace{-1em}
\resizebox{\textwidth}{!}{
\begin{tabular}{l|cccc|cccc|cccc}
\toprule
\multirow{2}{*}{Method} & \multicolumn{4}{c|}{\benchname-RestBench} & \multicolumn{4}{c|}{\benchname-MetaTool} & \multicolumn{4}{c}{\benchname-ToolBench} \\
\cmidrule{2-13}
& S@5 & S@10 & N@5 & N@10    & S@5 & S@10 & N@5 & N@10    & S@5 & S@10 & N@5 & N@10 \\
\midrule
Random Guess          & 0.00 & 2.59 & 10.78 & 16.09 & 3.70 & 5.93 & 1.99 & 2.56 & 0.00 & 0.00 & 0.18 & 0.58 \\
BM25                 & 6.67 & 17.78 & 34.50 & 38.32 & 37.78 & 47.78 & 30.94 & 33.63 & 1.11 & 12.22 & 47.93 & 51.13 \\ \midrule
BERT-base (ZS)            & 1.11 & 3.33 & 19.12 & 23.96 & 43.33 & 58.89 & 36.49 & 41.37 & 0.00 & 1.11 & 23.16 & 25.78 \\
RoBERTa-base (ZS)          & 1.11 & 3.33 & 21.59 & 28.15 & 41.11 & 44.44 & 35.41 & 35.69 & 0.00 & 0.00 & 6.04 & 6.40 \\
all-miniLM-L6-v2 (ZS)       & 15.56 & \underline{34.44} & \underline{52.55} & \underline{52.76} & \underline{81.11} & \underline{84.44} & \underline{71.99} & \underline{71.65} & 24.44 & 44.44 & \underline{67.16} & 65.33 \\
\midrule
BERT-base (FT)        & 3.33 & 8.89 & 30.00 & 35.55 & 45.56 & 56.67 & 37.58 & 40.50 & 5.56 & 10.00 & 38.03 & 40.59 \\
RoBERTa-base  (FT)    & 1.11 & 6.67 & 22.70 & 31.69 & 47.78 & 53.33 & 40.94 & 41.88 & 2.22 & 6.67 & 23.22 & 26.36 \\
all-miniLM-L6-v2 (FT)  & \underline{16.67} & 32.22 & 51.77 & 51.29 & \underline{81.11} & \underline{84.44} & \underline{71.99} & \underline{71.65} & \underline{28.89} & \textbf{56.67} & 67.02 & \textbf{66.39} \\ \midrule
QTA (Ours)             & \textbf{32.22} & \textbf{55.56} & \textbf{63.50} & \textbf{62.98} & \textbf{83.31} & \textbf{85.56} & \textbf{72.01} & \textbf{71.71} & \textbf{33.33} & \underline{46.67} & \textbf{67.81} & \underline{65.61} \\
\bottomrule
\end{tabular}}
\vspace{-1em}
\end{table*}

The above process is based on a single annotated sample, resulting in one DPO sample. 
Furthermore, we can instruct the LLM to execute the DPO sample generation process $m$ times on the same annotated sample, creating $m$ DPO samples for data augmentation. 
Subsequently, performing the aforementioned DPO training will enable the LLM to achieve robust alignment capabilities.
Within our tool retrieval system, the adopted approach enables dynamic adjustments to our query rewriting strategies. 
It continuously learns from user query rankings to refine outcomes. 
This method of strategy optimization not only diminishes the dependency on extensive gold standard datasets but also boosts the model's adaptability and precision across new domains and complex query scenarios.

\subsection{Inference}

As illustrated in~\cref{fig:framework}, after training the LLM, we perform inference on unseen samples.
Specifically, according to~\cref{eq:main_sub}, we prompt the LLM to rewrite user queries based on a few tool documents. 
Subsequently, the retrieval model utilizes these re-written queries to search the tool database, retrieving several relevant tools.

\section{Experiments}

\subsection{Settings}

\nbf{Baselines.}
We select two categories of baselines for our study, encompassing both non-deep learning and deep learning methods. 
Specifically, we employed the widely utilized Okapi BM25 model~\cite{10.1561/1500000019@bm25}, a standard non-neural information retrieval model. We used the standard implementation with default parameters (k1=$1.2$, b=$0.75$). This established model is based on term frequency and inverse document frequency metrics, and remains a benchmark choice in information retrieval tasks
Additionally, we implement a ``Random Guess'' strategy, which randomly selects tools from the database as retrieval results.
For deep learning methods, we include the following three popular methods
1) BERT-base-uncased~\cite{DBLP:conf/naacl/DevlinCLT19@bert}: a pioneering model in the NLP field, employing a base version of the bidirectional encoder representations model;
2) RoBERTa-base~\cite{DBLP:journals/corr/abs-1907-11692@roberta}: an optimized version of BERT with modifications for more robust performance;
3) all-MiniLM-L6-v2~\cite{DBLP:conf/nips/WangW0B0020@minilm}: a model specifically fine-tuned on over $1$ billion sentence pairs, designed for efficient sentence semantic tasks.
Additionally, we apply zero-shot (ZS) and fine-tuning (FT) strategies to these deep learning models.

\nbf{Implementation details.}
For fine-tuned models, we follow the procedures outlined in ToolBench~\cite{DBLP:journals/corr/abs-2307-16789@toolbench}, utilizing a similar dataset construction. 
Specifically, queries are treated as inputs, and the correct tool documents are used as the target to conduct the training.
In our \methodname framework, we utilize the all-MiniLM-L6-v2 as the retrieval model and Mistral-7B~\cite{mistralai@7B} as the LLM. 
In~\cref{eq:main_sub}, we randomly sample $5$ tools to form $T_{sub}$.
In the step for generating DPO samples, we define $n=10$ in~\cref{eq:ranking-function}.
Moreover, we generate $m=100$ DPO samples for DPO training for each annotated sample.
The DPO training is conducted with a batch size of $32$ over three epochs, utilizing a learning rate of $5e-6$.
We align our hyper-parameter settings in~\cref{eq:dpo-loss} with those established in prior research~\cite{DBLP:conf/nips/RafailovSMMEF23@dpo}.
All experiments are conducted by using an NVIDIA RTX $6000$ Ada with $48$ GB. 

\begin{table*}[!tp]
\centering
\caption{Visualization results: user query before and after rewriting in \benchname-RestBench dataset.}
\vspace{-1em}
\label{tab:qualitative}
\setlength{\tabcolsep}{5pt}
\begin{tabular}{l|m{8cm}|m{4.5cm}|c|c}
\toprule
 & User Query & \centering Golden Tools  & S@5 & N@5  \\
\midrule
Raw & I'm watching the tv series The Last Of Us and I need some more recommendations & \multirow{2}{4.5cm}[-0.75em]{\centering\begin{tabular}{@{}c@{}}
``GET /search/tv'' \\
``GET /tv/\{tv\_id\}/recommendations''
\end{tabular}} & 0 & 48.71 \\
\cmidrule{1-2} \cmidrule{4-5}
\begin{tabular}[c]{@{}l@{}}Re-written\end{tabular} & Search for TV show The Last Of Us to get its ID, then use that ID to get TV show recommendations & & 1 & 76.93 \\
\midrule
Raw & Avatar versus Avatar: The Way of Water, which has a higher rating & \multirow{2}{4.5cm}[-1.25em]{\centering\begin{tabular}{@{}c@{}}
``GET /search/movie'' \\
``GET /search/movie''
\end{tabular}} & 0 & 0.00 \\
\cmidrule{1-2} \cmidrule{4-5}
\begin{tabular}[c]{@{}l@{}}Re-written\end{tabular} & Search for movie Avatar and get its rating, then search for movie Avatar: The Way of Water and get its rating. Compare the ratings to determine which has a higher rating. & & 1 & 48.71 \\
\midrule
Raw & What is the genre of the movie Lord of the Ring? & \multirow{2}{4.5cm}[-0.7em]{\centering\begin{tabular}{@{}c@{}}
``GET /search/movie'' \\
``GET /movie/\{movie\_id\}''
\end{tabular}} & 0 & 0.00 \\
\cmidrule{1-2} \cmidrule{4-5}
\begin{tabular}[c]{@{}l@{}}Re-written\end{tabular} & Search for movie Lord of the Rings to get its ID, then retrieve the movie's details including its genre & & 1 & 76.93 \\
\bottomrule
\end{tabular}
\vspace{-1em}
\end{table*}

\subsection{Main Results}

As shown in~\cref{tab:main_results}, we compare various methods on \benchname benchmark across three sub-task: \benchname-RestBench, \benchname-ToolBench, and \benchname-MetaTool.
In summary, the proposed \methodname framework achieves either the best or the second-best results across all metrics.

In the \benchname-RestBench sub-task, our \methodname framework excels across all metrics. 
For example, compared to the suboptimal method of all-MiniLM-L6-v2 (FT), the \methodname achieves a $93.28\%$ improvement in $S@5$.
Although all-MiniLM-L6-v2 $+$ FT shows a $7.13\%$ increase in $S@5$ over its ZS result, it experiences declines in the metrics of $S@10$, $N@5$, and $N@10$.
The \benchname-RestBench task focuses on IMDB-related movie user queries and tools.
Many queries appear to require only one tool; however, IMDB's tool functionalities are finely divided, posing challenges for retrieval system to capture accurately. 
Our \methodname method enhances performance by logically rewriting user queries based on an analysis of partial tool examples.
Furthermore, we provide detailed examples and analysis in~\cref{ss:qualitative}.

In the \benchname-MetaTool sub-task, our \methodname framework achieves the highest scores. 
We observe that fine-tuning the all-miniLM-L6-v2 model shows no improvement in scores.
In MetaTool, the $199$ tools cataloged exhibit diversity but low similarity. 
Moreover, each query corresponds uniquely to one golden tool, simplifying the differentiation of each sample.
Therefore, the keyword-based BM25 method achieves a $S@5$ score of $37.78$. 
Moreover, the all-miniLM-L6-v2 model consistently achieves high scores, with $S@5$ and $S@10$ both surpassing $80$ points.
Therefore, we hypothesize that the pre-training dataset of the all-miniLM-L6-v2 contains numerous similar instances.
It has reached a saturation point, impeding further improvements and resulting in a performance plateau. 
Conversely, our \methodname framework achieves better performance by breaking aforementioned bottleneck, affirming its robustness.

In the \benchname-ToolBench subtask, our \methodname framework achieves the highest performance in top-$5$ retrieval, as shown in its $S@5$ and $N@5$ scores.
An analysis of ToolBench's sample characteristics reveals that although approximately $2,000$ tools are cataloged, user queries primarily involve only about $10$ tools. 
Specifically, user queries are predominantly focused on tools related to videos and television programs.
This similarity in usage patterns leads to consistent distributions across the training and test sets.
Consequently, retrieval models are capable of achieving enhancements even with a few training samples. 
Additionally, our \methodname framework, which integrates LLMs, shows improvements with reduced susceptibility to overfitting, implying that further performance gains could be realized with an increased sample size.

In summary, our \methodname framework shows SOTA performance well across all sub-tasks, showcasing its robust capabilities in \benchname benchmark. 
This indicates that the \methodname's advantage in complex multi-tools environments, enhancing retrieval performance.

\begin{table}[tp]
\centering
\caption{Ablation study on the impact of different modules within the LLM and Retrieval models using the \benchname-Restbench dataset. The results are shown in percentages. The best performance scores are highlighted in \textbf{bold}.}
\label{tab:ablation_modules}
\vspace{-1em}
\begin{tabular}{l|cccc}
\toprule
\multirow{2}{*}{Method} & \multicolumn{4}{c}{\benchname-RestBench} \\
\cmidrule{2-5}
& S@5 & S@10 & N@5 & N@10  \\
\midrule
\multicolumn{5}{c}{Retrieval Model Only} \\ \midrule
BM25            & 6.67 & 17.78 & 34.50 & 38.32 \\
all-MiniLM-L6-v2 (ZS)   & 15.56 & 34.44 & 52.55 & 52.76 \\ \midrule
\multicolumn{5}{c}{Fixed LLM+ Retrieval Model} \\
\midrule
QTA - BM25            & 17.78 & 42.22 & 56.55 & 52.94 \\
QTA - all-MiniLM-L6-v2   & \textbf{32.22} & \textbf{55.56} & \textbf{63.50} & \textbf{62.98} \\ \midrule
\multicolumn{5}{c}{LLM + Fixed Retrieval Model} \\
\midrule
QTA - Llama-3-8B               & 27.78 & 51.11 & 60.31 & 58.88  \\
QTA - Mistral-7B         & \textbf{32.22} & \textbf{55.56} & \textbf{63.50} & \textbf{62.98} \\
\bottomrule
\end{tabular}
\vspace{-1em}
\end{table}

\subsection{Qualitative Results}
\label{ss:qualitative}

In this section, the performance of our \methodname framework on the \benchname-RestBench dataset is visualized through the analysis of three randomly selected samples. 
As shown in~\cref{tab:qualitative}, we present the before and after versions of three sets of user queries, detailing the changes in the $S@5$ and $N@5$ metrics.
In summary, the \methodname framework generates high-quality queries, resulting in improvement over retrieval models. 
For instance, the initial query ``I'm watching the TV series The Last Of Us and I need some more recommendations'' could not accurately match golden tools, with $S@5$ being $0$. 
In contrast, the re-written query ``Search for TV show The Last Of Us to get its ID, then use that ID to get TV show recommendations'' successfully matches two golden tools, boosting $S@5$ from $0$ to $1$ and $N@5$ from $48.71$ to $76.93$. 
In the second example, the original query is too brief and cannot match any golden tools, resulting in both $S@5$ and $N@5$ being $0$. 
Following augmentation by the LLM, the query is re-written to ``Search for the movie Avatar and get its rating, then search for the movie Avatar: The Way of Water and get its rating. Compare the ratings to determine which has a higher rating.'', which presents more details than the original one.
This detailed query better aligns with the expectations of the retrieval model, facilitating effective tool matching.
Overall, the \methodname method addresses the shortcomings of previous retrieval systems when dealing with complex, multi-step tasks by intelligently rewriting user queries. 
The re-written queries are not only more specific and clear but also effectively break down the task into manageable steps, ensuring each step aligns with the correct tools. 

\subsection{Ablation Study}

In this section, we investigate the effects of various modules and configurations on the \methodname framework and further explore its generalizability.

\begin{table}[tp]
\centering
\caption{Impact of training sample quantity on performance. ``ASam'' indicates the number of annotated samples and ``TSam'' means the number of training samples. The $m$ presents the number of iterations for generating DPO samples as detailed in~\cref{ss:training}. The results are shown in percentages. The best results are in \textbf{bold}.}
\label{tab:train_samples}
\vspace{-1em}
\resizebox{0.48\textwidth}{!}{
\begin{tabular}{l|lll|cccc}
\toprule
\multirow{2}{*}{Method} & \multirow{2}{*}{ASam} & \multirow{2}{*}{$m$} & \multirow{2}{*}{TSam} & \multicolumn{4}{c}{\benchname-RestBench} \\
\cmidrule{5-8}
& & & & S@5 & S@10 & N@5 & N@10  \\
\midrule
Baseline        & - & - & - & 15.56 & 34.44 & 52.55 & 52.76 \\
\midrule
             & 1 & $\times 1$ & 1 & 20.00 & 42.22 & 56.67 & 53.78  \\
QTA             & 1 & $\times 10$ & 10& 24.44 & 44.44 & 51.11 & 55.91  \\
             & 1 & $\times 100$ & 100 & 27.78 & 46.67 & 60.43 & 57.06  \\
\midrule
             & 10 & $\times 1$ & 10 & 25.56 & 44.44 & 60.31 & 58.88   \\
QTA             & 10 & $\times 10$ & 100 & 28.89 & 44.44 & 60.98 & 60.24   \\
             & 10 & $\times 100$ & 1000 & \textbf{32.22} & \textbf{55.56} & \textbf{63.50} & \textbf{62.98}  \\
\bottomrule
\end{tabular}}
\vspace{-1em}
\end{table}

\begin{table*}[tp]
\centering
\caption{Comparative analysis of model generalizability across \benchname-RestBench and \benchname-ToolBench dataset, with the best score \textbf{bolded}. The results are shown in percentages. All methods are trained only in the \benchname-RestBench dataset.}
\label{tab:cross-dataset}
\vspace{-1em}
\begin{tabular}{l|cccc|cccc}
\toprule
\multirow{2}{*}{Method} & \multicolumn{4}{c|}{\benchname-RestBench} & \multicolumn{4}{c}{\benchname-ToolBench} \\
\cmidrule{2-9}
 & S@5 & S@10 & N@5 & N@10 & S@5 & S@10 & N@5 & N@10  \\
\midrule
Random Guess          & 0.00 & 2.59 & 10.78 & 16.09 & 0.00 & 0.00 & 0.18 & 0.58 \\ \midrule
BERT-base        & 3.33 & 8.89 & 30.00 & 35.55 & 0.00 & 1.11 & 24.81 & 25.04 \\
RoBERTa-base     & 1.11 & 6.67 & 22.70 & 31.69 & 0.00 & 0.00 & 7.72 & 9.49 \\
all-miniLM-L6-v2 & 16.67 & 32.22 & 51.77 & 51.29 & 24.44 & 40.00 & 67.29 & 63.96 \\ \midrule
QTA               & \textbf{32.22} & \textbf{55.56} & \textbf{63.50} & \textbf{62.98} & \textbf{30.00} & \textbf{52.22} & \textbf{67.79} & \textbf{64.49}  \\
\bottomrule
\end{tabular}
\vspace{-1em}
\end{table*}

\subsubsection{Ablating Main Modules}

To further explore the impact of different components on the performance of the \methodname framework, we conduct detailed ablation experiments, as shown in~\cref{tab:ablation_modules}. 
Specifically, we explore the effects of LLM and retrieval models under fixed and variable configurations.
Initially, we explore the impact of different retrieval models on the \methodname framework under a fixed LLM. 
The findings indicate that the implementation of our framework leads to an overall enhancement in the performance of retrieval systems. 
This demonstrates the effectiveness of aligning queries with tool document descriptions. 
Furthermore, the results consistently show that all-MiniLM-L6-v2 outperforms BM25 across all metrics, and this trend is maintained under the QTA framework.
Our framework substantially boosts the performance of weaker retrieval models. 
For instance, QTA-BM25 improves from $34.50$ to $56.55$ in $N@5$, surpassing even the performance of all-MiniLM-L6-v2.
After that, we examine the impact of different LLMs on the \methodname method under a fixed retrieval model. 
The experimental results show that when using the Mistral-7B model, the \methodname achieves the best performance across all metrics, outperforming Llama-3-8B comprehensively. 
Despite some studies indicating that Mistral-7B falls short of Llama-3-8B in various tasks such as MMLU~\cite{DBLP:conf/iclr/HendrycksBBZMSS21@mmlu}, HumanEVal~\cite{DBLP:journals/corr/abs-2107-03374@humaneval}, and MATH~\cite{DBLP:conf/nips/HendrycksBKABTS21@math}, it excels in instruction-following rewriting tasks as shown in~\cref{tab:ablation_modules}. 
This discrepancy indicates that our task design and process are more challenging. 
Therefore, we suggest that users and researchers prioritize assessing the reasoning capabilities of the LLM over its commonsense understanding.

Through ablation experiments with different combinations of LLMs and retrieval models, we find that the realization of optimal performance depends on the effective combination of LLM and retrieval models. 
Using advanced LLMs and retrieval models can complement each other, enhancing the system's understanding of user queries and matching golden tools.

\subsubsection{The Number of Training Samples}

As shown in~\cref{tab:train_samples}, to assess the effectiveness of our proposed \methodname framework under low-resource conditions, we report the results with varying numbers of training samples.
The ``Baseline'' method is the backbone retrieval model ``all-MiniLM-L6-v2 (ZS)''.
The results show that even with just one training sample, the \methodname outperforms the baseline across all metrics. 
Specifically, the \methodname with $1$ ``TSam'' boosts $S@5$ from $15.56$ to $20.00$ on the baseline, demonstrating the effectiveness of the QTA method in extremely low-resource scenarios. 
As the number of $m$ increases to $10$ and $100$, its performance keeps improving. 
This indicates that even with very sparse annotated data, the \methodname method can improve model performance by increasing the number of samples through Ranking-based DPO data synthesis.
Furthermore, our observations indicate that our framework exhibits comparable performance when the quantity of TSam remains consistent ($10$, $100$), despite variations in annotated samples. 
This suggests that our approach is robust and functions effectively even with limited manual annotations.

\subsubsection{Cross Task Generalizability}

As shown in~\cref{tab:cross-dataset}, we explore the generalization capabilities of various methods across different subsets.
In detail, the models are trained on the training set of \benchname-RestBench dataset and then evaluated on the test set of \benchname-RestBench and \benchname-ToolBench datasets.
For zero-shot evaluation on the \benchname-ToolBench dataset, our \methodname framework achieves the SOTA performance.
This indicates that the \methodname has strong generalization capabilities on unseen datasets.
In comparison, the all-miniLM-L6-v2 method performs second best on the \benchname-ToolBench dataset. 
Other baseline methods like BERT-base and RoBERTa-base perform poorly in terms of generalization, especially on the \benchname-ToolBench dataset, with $S@5$ and $S@10$ scores nearly zero.
The performance of these models closely approximates ``Random Guess'' on the $S@k$ metric, indicative of their near inability to transfer knowledge across datasets.
The cross-dataset experimental results confirm the wide applicability and efficiency of the \methodname method in different task scenarios. 

\section{Conclusion}

This paper introduces a new \benchname benchmark for evaluating MTR tasks under real-world tool-augmented LLM scenarios with a large number of tools. 
This benchmark consists of three subsets, each providing $90$ test samples and $10$ training samples. 
To address the MTR task, we proposed the \methodname framework, which aligns user queries with tool documents under low-resource conditions by leveraging ranking functions alongside DPO training.
Our experimental results demonstrated that the \methodname framework significantly outperforms traditional baselines across most subsets, especially in low-resource environments, underscoring its effectiveness in improving retrieval performance.
Our ablation study highlights the robustness of \methodname, showing that it performs well even with minimal training data through effective data synthesis strategies. 
Additionally, \methodname exhibited strong cross-dataset generalizability, maintaining high performance when applied to unseen datasets, which underscores its potential for real-world applications.

{
\bibliographystyle{ACM-Reference-Format}
\bibliography{sample-base, custom}
}

\end{document}